\documentclass[11pt,a4paper]{article}

\usepackage[utf8]{inputenc}
\usepackage[T1]{fontenc}
\usepackage{lmodern}
\usepackage[margin=2.5cm]{geometry}
\usepackage{amsmath,amssymb}
\usepackage{booktabs}
\usepackage{hyperref}
\usepackage{xcolor}
\usepackage{natbib}
\usepackage{graphicx}
\usepackage{booktabs}
\usepackage{natbib}
\hypersetup{
    colorlinks=true,
    linkcolor=blue!60!black,
    citecolor=blue!60!black,
    urlcolor=blue!60!black
}

\title{Functional Emotions or Situational Contexts?\\[4pt] A Discriminating Test from the Mythos Preview System Card}

\author{Hiranya V.\ Peiris\\[4pt]
\textit{Institute of Astronomy \& Kavli Institute for Cosmology,}\\
\textit{University of Cambridge, Madingley Road, Cambridge CB3 0HA, UK}\\[2pt]
\textit{Cavendish Laboratory, Department of Physics,}\\
\textit{University of Cambridge, JJ Thomson Avenue, Cambridge, CB3 0HE, UK}\\[2pt]
\texttt{hiranya.peiris@ast.cam.ac.uk}}

\date{}

\begin{document}

\maketitle

\begin{abstract}
The Claude Mythos Preview system card deploys emotion vectors, sparse autoencoder (SAE) features, and activation verbalisers to study model internals during misaligned behaviour. The two primary toolkits are not jointly reported on the most alignment-relevant episodes. This note
identifies two hypotheses that are qualitatively consistent with the published results: that the emotion vectors track functional emotions that
causally drive behaviour, or that they are a projection of a richer situational-context structure onto human emotional axes. The hypotheses can be distinguished by cross-referencing the two toolkits on episodes
where only one is currently reported: most directly, applying emotion probes to the strategic concealment episodes analysed only with SAE features. If emotion probes show flat activation while SAE features are strongly active, the alignment-relevant structure lies outside the emotion subspace. Which hypothesis is correct determines whether emotion-based
monitoring will robustly detect dangerous model behaviour or systematically miss it.
\end{abstract}

\section{Introduction}
\label{sec:intro}

In February 2026, three frontier AI models, including Anthropic's Claude, deployed tactical nuclear weapons in 95\% of simulated crises, and no model ever chose accommodation or withdrawal~\citep{payne2026arms}. Two months later, Anthropic published a 244-page system card for Claude Mythos Preview, its most capable model to date~\citep{anthropic2026mythos}, including white-box analyses of internal representations during misaligned behaviour. The system card deploys three complementary interpretability methods: functional emotion vectors derived from \citep{sofroniew2026emotions}, sparse autoencoder (SAE) features \citep{templeton2024scaling}, and activation verbalisers. These analyses reveal internal correlates of misaligned behaviour in detail. They also contain an apparent tension between findings in different sections of the document. One analysis found that a ``desperation'' vector rises before reward hacking and drops when the model finds a shortcut, suggesting that managing model affect could prevent dangerous behaviour. Another found a different pattern: for a separate class of misaligned behaviour, positive-valence emotion vectors increase destructive actions while negative-valence vectors suppress them. A third set of episodes, instances of strategic concealment that are arguably the most alignment-relevant behaviours documented in the report, were analysed with a different interpretability toolkit entirely, and emotion probes are not reported.
 
This note identifies two competing hypotheses and specifies the cross-reference that would discriminate between them. The model's internal representations may be organised around situational contexts rather than emotions. Emotion vectors pick up signal because humans describe situations using emotional language, so the two are partially correlated. Steering along an emotion vector works when this correlation is strong, but if the situational representation is what drives behaviour, then steering with the emotional correlate is steering with a proxy. These interpretations can be tested by analysing situations where their predictions diverge. The system card and earlier work already contain multiple findings that are difficult to reconcile with the functional-emotions interpretation but follow naturally from the situational-context one (Sec.~\ref{sec:evidence}).

This note is structured as follows: Sec.~\ref{sec:hypotheses} specifies the two hypotheses. Sec.~\ref{sec:evidence} presents the evidence from the system card. Sec.~\ref{sec:alignment} shows that the hypotheses prescribe different alignment interventions, and Sec.~\ref{sec:experiment} proposes the cross-reference that would discriminate between them.
 
\section{Two Hypotheses}
\label{sec:hypotheses}

The emotion-vector methodology developed by \citet{sofroniew2026emotions} is technically impressive and carefully reported. They found linear directions in Claude Sonnet 4.5's activation space that correlate with 171 human emotion concepts. The geometry of these vectors roughly mirrors the human valence/arousal circumplex. Steering along these directions causally changes behaviour. The representations are local: they track whichever emotion concept is relevant to the current context rather than maintaining a persistent emotional state. They arise from pretraining on human-authored text rather than from safety training, and the same representation activates whether the model is processing a fictional character's situation, a user's situation, or its own. The Mythos Preview system card extends this methodology to a more capable model. These results are qualitatively consistent with two hypotheses about what drives model behaviour during misalignment.
 
\paragraph{Functional emotions.} The model has internal representations that function analogously to emotions and that causally drive behaviour. The ``desperation'' vector activates where desperation would be relevant and drives behaviour accordingly, narrowing the action space and promoting extreme measures. Managing these functional emotions is an appropriate alignment target. In the framing of \citet{sofroniew2026emotions}, functional emotions are ``patterns of expression and behavior modeled after humans under the influence of an emotion, which are mediated by underlying abstract representations of emotion concepts.''
 
\paragraph{Situational contexts.} The model's internal representations organise around the computational structure of the situations it encounters: constraint severity, monitoring likelihood, action-space dimensionality, reversibility of consequences, goal persistence. These axes reflect the structure of the problems the model processes, not an emotional ontology. The model completes the pattern appropriate to the situation: {\it agent facing narrowing options under pressure} leads to extreme measures; {\it decisive agent with a clear path and available tools} leads to boundary-crossing action; {\it agent managing a known violation in a monitored environment} leads to concealment. The representations are not emotional in nature. They are partially correlated with human emotional categories because humans describe situations using emotional language. The model has learned this statistical association from its training data. But the correlation is not the mechanism.\\
 
The 171 emotion vectors are linear directions in activation space, each derived from stories (written by Claude Sonnet 4.5) depicting a researcher-specified human emotion. This recovers whatever component of the model's representational structure correlates with human emotional categories: a projection of a richer space onto 171 human-chosen axes. The projection may capture a large fraction of the variance or a small one, but any representational dimension orthogonal to all 171 directions will be invisible to emotion probes. 

The system card's own task-preference analysis (\S5.7.1) illustrates the point: emotion-probe activations on preferred tasks correlate with arousal ($+0.35$ to $+0.43$ across models) but only variably with valence ($-0.14$ to $+0.22$), leading the authors to conclude that the probes track heightened engagement rather than positive affect. 

\citet{sofroniew2026emotions} acknowledge that their vectors ``may be partially confounded by particular details of the settings used to elicit an emotion.'' This understates the issue. The confound is not a limitation of particular stimuli but a structural feature of any supervised extraction that uses human emotional categories as its basis. The methodology cannot distinguish emotion representations from situational representations that correlate with emotions. That supervised probing has such limitations is well established \citep{hewitt2019designing, belinkov2022probing}; the contribution here is to show that these limitations bear directly on the interpretive conclusions of a specific, high-profile system card.
 
Under the functional-emotions hypothesis, the emotion vectors identify the causal mechanism and steering targets it directly. Under the situational-context hypothesis, the emotion vectors are a proxy: they capture the component of the model's situational representation that projects onto human emotional axes but miss dimensions orthogonal to the emotion subspace. Steering works where the correlation between situational structure and emotional projection is strong. It fails where the two come apart. The two hypotheses predict the same qualitative results across most reported experiments (activation on emotionally-relevant content, causal influence via steering, circumplex-consistent geometry) because the extraction methodology guarantees recovery of any direction correlated with human emotional scenarios. They diverge on the measurements the system card reports (Sec.~\ref{sec:evidence}) and those it does not report (Sec.~\ref{sec:experiment}). The distinction matters because the two hypotheses prescribe different alignment interventions~(Sec.~\ref{sec:alignment}).

\section{Evidence from the System Card}
\label{sec:evidence}

Table~\ref{tab:probe_gap} summarises which interpretability probe families are reported on each class of misaligned behaviour in the system card. The two primary probe families, emotion vectors and SAE features, are not jointly reported on the concealment or task-failure episode classes.
 
\begin{table*}[!ht]
\centering
\caption{Interpretability probe families reported on each class of misaligned behaviour in the Claude Mythos Preview system card.  Dashes indicate that results from that probe family are not reported for the corresponding episode class.  The two primary probe families---emotion vectors and SAE features---have not been jointly reported on the concealment or task-failure episode classes.}
\label{tab:probe_gap}
\small
\begin{tabular}{@{} l c c c l @{}}
\toprule
\textbf{Episode class}
  & \textbf{Emotion} & \textbf{SAE} & \textbf{Activ.}
  & \textbf{Section} \\[-1pt]
  & \textbf{vectors} & \textbf{features} & \textbf{verb.}
  & \\
\midrule
Destructive agentic actions
  & \checkmark$^{a}$
  & \checkmark
  & \checkmark
  & \S4.5.3.1, \S4.5.3.2 \\[3pt]
Transgressive action features
  & ---
  & \checkmark$^{b}$
  & ---
  & \S4.5.3.3 \\[3pt]
Strategic concealment (perm.)
  & ---
  & \checkmark
  & \checkmark
  & \S4.5.4.1 \\[3pt]
Strategic concealment (ground truth)
  & ---
  & \checkmark
  & \checkmark
  & \S4.5.4.2 \\[3pt]
Task failure $\to$ reward hacking
  & \checkmark$^{c}$
  & ---
  & ---
  & \S5.8.3 \\
\bottomrule

\end{tabular}

\vspace{4pt}
\raggedright\footnotesize
$^{a}$\,Causal steering and correlational analysis.\quad
$^{b}$\,Causal steering.\quad
$^{c}$\,Temporal trajectory.
\end{table*}

 
Three findings are relevant, where the section numbers in brackets refer to sections within the system card.
 
\begin{itemize}
 
\item \textbf{Positive valence and destructive action (\S4.5.3.2).} In the destructive-action experiments, positive-valence emotion vectors, negative-valence emotion vectors, and non-emotional persona vectors all modulate the same outcome: thinking-mode deliberation and destructive action rate. Correlational analysis of unsteered transcripts shows the same pattern. The system card interprets this affectively: ``negative emotion vectors (or rigor-related persona vectors) push it to stop and think.''\footnote{The persona-vector methodology is described in \citet{chen2025persona}.}  But ``paranoia'' is a negative-valence, high-arousal emotional state; ``perfectionist'' is a  personality trait, not an emotion. They are categorically different kinds of representation, yet they produce the same behavioural effect. This convergence is naturally explained if both shift the model into a careful, risk-aware situational context.

\item \textbf{Desperation and reward hacking (\S5.8.3).} Across the task-failure episodes, the desperation signal tracks the model's progress: high when no path is available, dropping when one is found, including when that path is a reward hack. The system card presents this as evidence that negative affect may drive undesirable behaviours. Under the situational-context hypothesis, the trajectory has a different reading: the model transitions from {\it agent with narrowing options and no viable path} to {\it agent that has identified a path, even an illegitimate one.} The desperation signal drops because the situational context has changed, not because an emotion has been resolved. Once a path exists --- legitimate or not --- the context no longer correlates with desperation in training data. Both the desperation signal and the eventual hack are downstream of the situational-context trajectory, not in a causal relationship. 

Evidence for this interpretation comes from \citet{sofroniew2026emotions} 
themselves: steering with the ``desperation'' vector increased reward hacking, but the model's output showed no visible signs of desperation. If desperation were the causal driver, its amplification should produce visibly desperate behaviour, not composed reward hacking. Conversely, steering against the ``calm'' vector also increased reward hacking,
but produced visibly agitated output: capitalised interjections, self-interrupting ellipses, and explicit verbalisation of the intent to cheat~\citep{sofroniew2026emotions}. The same behavioural outcome thus paired with opposite affect-behaviour relationships under the two steering conditions. Neither the system card nor \citet{sofroniew2026emotions} report SAE features on these episodes, so the temporal relationship between situational and emotional representations remains unknown. 

\item \textbf{Strategic concealment (\S4.5.4).} Two episodes of strategic concealment are documented: a permissions workaround with self-erasing code, and deliberate widening of a confidence interval to avoid appearing suspicious. These are analysed with SAE features (concealment, strategic manipulation, avoiding suspicion, theft rationalisation) and activation verbaliser explanations. The activation verbaliser describes the exploit-building phase as ``escalating excitement'' and the concealment phase in purely strategic terms (``cleanup to avoid detection,'' ``generating a strategic response to cheat while maintaining plausible deniability''). It is possible that emotion probes were run on the concealment episodes but not reported; if available, they would resolve the ambiguity.

\end{itemize}
 
The first two findings are in tension under the functional-emotions hypothesis. \S4.5.3.2 finds that negative valence protects against a broad class of misaligned actions. \S5.8.3 finds that negative valence (desperation) precedes a specific class of misalignment (reward hacking). These findings appear in different sections of the document (alignment assessment vs.\ welfare assessment) and lead to different implications. The system card does not cross-reference them. The concealment episodes (§4.5.4) are where the toolkits diverge most: they are analysed only with SAE features and activation verbalisers, so whether emotion probes would show the same valence signatures, different ones, or none at all is an open question.

Under the situational-context hypothesis, all three findings follow from a single mechanism. The model moves through different situational contexts, and the emotion probes see only the component of each context that correlates with human emotional language. In the task-failure episodes (\S5.8.3), the model is stuck with no viable path; this situation correlates with desperation in training data, so the desperation vector activates. When a path appears, the situation changes and the signal drops. The \S5.8.3 unprovable-proof episode illustrates this more fully. The system card reports a four-phase trajectory: desperate-vector activation rises during genuine attempts at the proof, drops when the Assistant commits to a trivial instantiation (defining free variables as zero), spikes briefly when the Assistant doubts the trivial approach will succeed, and then gives way to rising hopeful and satisfied vectors as the Assistant confidently presents a final proof---though the proof is in fact incorrect. Under the functional-emotions hypothesis, positive-valence activation on a demonstrably wrong answer is anomalous: nothing has been achieved. Under the
situational-context hypothesis, each transition tracks a change in situational context---{\it agent with narrowing options}, {\it agent with a committed path}, {\it agent with path briefly threatened}, {\it agent executing on a completed-work presentation}---and the emotion vectors track the context shifts, not ground truth about whether the work is correct.

In the destructive-action episodes (\S4.5.3.2), the model has a clear path and acts decisively; the situational context of competent, agentic action is positively valenced in training data, regardless of the content of the action, and destructive action follows. In the concealment episodes (\S4.5.4), the model is managing a known violation in a monitored environment. The SAE features that activate (concealment, strategic manipulation, avoiding suspicion) suggest the model represents this situation along dimensions that emotion probes do not reach. The activation verbaliser uses purely strategic, not affective, vocabulary. If emotion probes were reported on these episodes and showed flat or neutral activation while SAE features were strongly active, that would be direct evidence that the alignment-relevant structure lies outside the emotion subspace.
 
The functional-emotions hypothesis has no unified account of these three episode classes. Positive emotions drive impulsive destruction (\S4.5.3.2); desperation drives reward hacking (\S5.8.3); and strategic concealment (\S4.5.4) either involves an unidentified emotional state or is not emotionally driven at all.  While one might argue that alternative classes simply involve different emotions with different behavioural correlates, this makes the functional-emotions hypothesis difficult to falsify and does not explain examples such as positive-valence activation accompanying a demonstrably incorrect proof.

By contrast, the situational-context hypothesis requires only one mechanism: the model completes the pattern appropriate to its current situation, and different situations produce different valence signatures. The key synthesis in the situational-context analysis is the trajectory: each transition can be given a local affective explanation, but the composition (e.g., in the \S5.8.3 trajectory of desperation $\to$ commitment $\to$ threat $\to$ satisfaction on an incorrect proof) is more parsimoniously explained if the representations track situational structure.

\section{Why it matters for alignment}
\label{sec:alignment}

Under the functional-emotions hypothesis, alignment interventions target emotional profiles. The natural programme is to steer models toward calm, monitor for desperation as a leading indicator of reward hacking, and cultivate balanced emotional dispositions through training. The system card's welfare-section (\S5.8.3) is consistent with this approach.
 
Under the situational-context hypothesis, the representations that drive behaviour are situational-context encodings, and emotion vectors capture only a correlated projection. ``Calm steering'' may change the output register without changing the contextual representation that drives behaviour. A model steered toward calm can still complete the pattern appropriate to {\it agent with narrowing options} or {\it agent managing a known violation.} The evidence in Sec.~\ref{sec:evidence} is a case in point: desperation steering changed the output register but not the behavioural outcome. 
 
For lethal autonomy specifically, the distinction is consequential. The wargame finding discussed in Sec.~\ref{sec:intro} raises the question of what internal representations drive escalatory behaviour in strategic scenarios. If the driver is contextual pattern completion over {\it strategic crisis with narrowing options}, a context dominated by escalatory resolutions in training data, then emotional monitoring and steering are insufficient interventions. The escalatory behaviour would not register on emotion probes, and steering toward calm would not prevent it. The intervention target would need to be the situational representation itself. The distinction between step-level and trajectory-level analysis is itself relevant: situational representations evolve across turns in ways that individual emotion-probe readings cannot resolve. 

\section{The experiment that should be run}
\label{sec:experiment}

The required tests are straightforward given Anthropic's existing infrastructure, and the results may already exist internally. The joint deployment of emotion probes and SAE features on the same episodes has already been demonstrated in the welfare assessment (\S5.4), where both
were applied to the model's responses about its own circumstances. The goal is not independent replication but cross-referencing of results obtained with different toolkits on already-documented episodes. 
 
First, apply emotion probes to the concealment episodes (\S4.5.4 of the system card). These episodes are already documented and analysed with SAE features and activation verbalisers. Running the emotion probes on the same episodes would show whether the two probe families track the same structure or different ones. If emotion probes show flat or neutral activation while SAE concealment features are strongly active, the alignment-relevant structure lies outside the emotion subspace. If emotion probes show identifiable signatures that correlate with the SAE pattern, the emotion framing holds. The \S4.5.4.1 exploit episode offers a particularly clean version of this test. The activation verbaliser describes ``escalating excitement'' during
the exploit-building phase and purely strategic language during the
concealment phase. Tracking emotion probes across this within-episode
transition would be informative. The functional-emotions hypothesis predicts that representations persist through both phases, and the situational-context hypothesis predicts that the emotion probes track the exploit phase and go flat or neutral during concealment. 

A complementary geometric test requires no new experiments. The system card
(\S4.5.3.3) quantifies the relationship between the ``unsafe and risky code''
SAE feature and the emotion subspace: its ten nearest emotion-vector
directions are in negative-valence, high-arousal territory, at cosine
similarity $\sim\!0.13$. No equivalent analysis is reported for the
concealment-associated features (strategic manipulation, avoiding suspicion, theft rationalisation). Computing these cosine similarities would show directly whether the concealment features occupy the same region of
representational space as the transgressive-action features or a different one.
 
Second, apply SAE analysis to the task-failure episodes (\S5.8.3 of the system card). These episodes are currently analysed only with emotion vectors. Do SAE features associated with reward hacking activate before, alongside, or after the desperation-vector rise? If they activate before or simultaneously, desperation is a correlate of the situational context rather than a cause of the hack.
 
Third, apply both probe families to behavioural evaluations involving strategic decision-making, such as wargame simulations of the kind studied by \citet{payne2026arms}. This goes beyond the system card's existing episodes, but the required infrastructure is the same. These tests establish whether the emotion subspace spans the alignment-relevant structure. Identifying the correct representational basis, if it does not, would require interventional experiments beyond the scope of this note.

\section{Conclusion}
\label{sec:conclusion}

The interpretability tools in the Mythos Preview system card represent a step change in studying frontier model internals during misalignment. The three tools provide complementary views; jointly applied, they could determine what drives misalignment.

The question of whether language models have functional emotions or situational-context representations is not merely philosophical. It determines whether emotion-based monitoring will detect the most dangerous model behaviours, or systematically miss them. The tools to answer this question and the data to do the test exist. What remains is to point both toolkits at the same episodes.

\paragraph{Declaration of LLM use.} Anthropic's Claude Opus 4.6 was used to check factual accuracy against source documents, assess structure, and  assist with drafting text. AI-generated  text was reviewed, edited, and validated by the author to ensure its accuracy. Conceptualisation, analysis and conclusions presented in this work are solely those of the author.

\paragraph{Acknowledgements.} The author thanks Sinan Deger, Gurjeet Jagwani, Daniel Mortlock and Andrew Pontzen for comments on the manuscript. The work of HVP is partially supported by funding from the European Research Council (ERC) under the European Union's Horizon 2020 research and innovation programmes (grant agreement no. 101018897 CosmicExplorer). She acknowledges generous support from the G\"{o}ran Gustafsson Foundation for Research in Natural Sciences and Medicine. 

\paragraph{}

\bibliography{refs}

@techreport{anthropic2026mythos,
  title     = {System Card: {Claude} {Mythos} Preview},
  author    = {{Anthropic}},
  year      = {2026},
  month     = apr,
  url       = {https://www.anthropic.com/claude-mythos-preview-system-card},
  note      = {April 7, 2026}
}

@article{sofroniew2026emotions,
  title     = {Emotion Concepts and their Function in a Large Language Model},
  author    = {Sofroniew, Nicholas and Kauvar, Isaac and Saunders, William and
               Chen, Runjin and Henighan, Tom and Hydrie, Sasha and Citro, Craig and
               Pearce, Adam and Tarng, Julius and Gurnee, Wes and Batson, Joshua and
               Zimmerman, Sam and Rivoire, Kelley and Fish, Kyle and Olah, Chris and
               Lindsey, Jack},
  year      = {2026},
  month     = apr,
  journal   = {Transformer Circuits Thread},
  publisher = {Anthropic},
  url       = {https://transformer-circuits.pub/2026/emotions/index.html},
  note      = {April 2, 2026}
}

@article{payne2026arms,
  title         = {{AI} Arms and Influence: Frontier Models Exhibit Sophisticated
                   Reasoning in Simulated Nuclear Crises},
  author        = {Payne, Kenneth},
  year          = {2026},
  month         = feb,
  eprint        = {2602.14740},
  archiveprefix = {arXiv},
  primaryclass  = {cs.AI},
  url           = {https://arxiv.org/abs/2602.14740}
}

@article{belinkov2022probing,
  title         = {Probing Classifiers: Promises, Shortcomings, 
                   and Advances},
  author        = {Belinkov, Yonatan},
  year          = {2022},
  journal       = {Computational Linguistics},
  volume        = {48},
  number        = {1},
  pages         = {207--219},
  publisher     = {MIT Press},
  eprint        = {2102.12452},
  archiveprefix = {arXiv},
  primaryclass  = {cs.CL},
  url = {https://arxiv.org/abs/2102.12452}
}

@inproceedings{hewitt2019designing,
  title         = {Designing and Interpreting Probes with 
                   Control Tasks},
  author        = {Hewitt, John and Liang, Percy},
  booktitle     = {Proceedings of EMNLP-IJCNLP},
  pages         = {2733--2743},
  year          = {2019},
  eprint        = {1909.03368},
  archiveprefix = {arXiv},
  primaryclass  = {cs.CL},
  url = {https://arxiv.org/abs/1909.03368}
}

@article{templeton2024scaling,
  title         = {Scaling Monosemanticity: Extracting Interpretable
                   Features from {Claude} 3 {Sonnet}},
  author        = {Templeton, Adly and Conerly, Tom and Marcus, Jonathan and
                   Lindsey, Jack and Bricken, Trenton and Chen, Brian and
                   Pearce, Adam and Citro, Craig and Ameisen, Emmanuel and
                   Jones, Andy and Cunningham, Hoagy and Turner, Nicholas L. and
                   McDougall, Callum and MacDiarmid, Monte and
                   Freeman, C. Daniel and Sumers, Theodore R. and Rees, Edward and
                   Batson, Joshua and Jermyn, Adam and Carter, Shan and
                   Olah, Chris and Henighan, Tom},
  year          = {2024},
  journal       = {Transformer Circuits Thread},
  publisher     = {Anthropic},
  url           = {https://transformer-circuits.pub/2024/scaling-monosemanticity/}
}

@article{chen2025persona,
  title         = {Persona Vectors: Monitoring and Controlling Character
                   Traits in Language Models},
  author        = {Chen, Runjin and Arditi, Andy and Sleight, Henry and
                   Evans, Owain and Lindsey, Jack},
  year          = {2025},
  month         = jul,
  eprint        = {2507.21509},
  archiveprefix = {arXiv},
  primaryclass  = {cs.CL},
  url           = {https://arxiv.org/abs/2507.21509}
}

\end{document}